\documentstyle[preprint,aps]{revtex}

\begin{document}
\draft     
\title
{Repulsive Core  of NN S-Wave Scattering \linebreak
in a Quark Model with a Condensed Vacuum}
\author{P. J. A. Bicudo, L. S. Ferreira, C.M. Pl\'acido, J. E. F. T. Ribeiro }
\address{Departamento de F\'isica and CFIF-Edif\'icio Ciencia\\
 Instituto Superior T\'ecnico, Avenida Rovisco Pais,
1096 Lisboa Codex, Portugal}
\maketitle
\begin{abstract}
We work in a chiral invariant quark model, 
with a condensed vacuum, characterized by only one 
parameter. Bound state equations for the nucleon and $\Delta$
are solved in order to obtain an updated value of their radii and masses. 
Nucleon-nucleon S-Wave scattering is studied in the
RGM framework both for isospin T=1 and T=0.
The phase shifts are calculated and an equivalent local potential, which 
is consistent with K-N scattering, is derived.
The result is a reasonable microscopic short 
range repulsion in the nucleon-nucleon interaction.
\end{abstract}


\section{INTRODUCTION}

The NN central repulsion is a prominent experimental feature of 
NN scattering and through the years it has been the subject of many 
theoretical papers. The explanation to why this NN repulsion 
happens falls 
into two main classes: the skyrmion picture \cite{Skyrmion} and the 
resonating group method 
approach (RGM)\cite{Ribeiro,Oka,toki}. Both approaches yield a sizeable NN central 
repulsion and the question arises whether or not these models can also 
explain other aspects of the NN force and whether and if they can be brought 
to describe other hadronic phenomena. This extensive program 
is not completely met yet up to the present day . 
\par
An effective model\cite {Yaouanc,Modelo} 
has been developped in a series of papers,  with just one scale 
besides the quark mass parameter, which amounts to a BCS theory of 
``Cooper-like'' quark antiquark pairs with the quantum numbers of 
the vacuum: $^3P_0$. 
 This model embodies in a natural way the physics of spontaneous 
chiral symmetry breaking $S\chi SB$ and therefore has a ``correct'' chiral 
limit so that it is expected to fare well in describing low energy 
phenomenology. In our model the mechanism 
of $S\chi SB$  is triggered by the effective quark potential whose strength 
 besides governing the extent of the quark condensation also furnishes a 
natural scale (the typical hadronic size) which turns out to be crucial to 
understand exotic scattering cross sections\cite{KN}. In this way we were 
able to describe a considerable number of experimental results ranging from 
hadronic spectrum and exotic scattering cross sections to the coupling of 
pions to nucleons and deltas \cite{fpindelta}. 
As for the main weak point of 
this model we must refer that it yields a too small $f_\pi$. This is a 
consequence of lack of covariance\cite{covariance} together with the absence 
of coupled channels in the BCS treatment of this many-body problem.
For recent developments on spontaneous chiral symmetry breaking in quark 
models, see \cite{Szczepaniak}. 
\par
So it is interesting to see whether we can obtain a reasonable $NN$ repulsion 
with the same potential strength  used in the above cited papers. The key 
point is that the physics of the $NN$ central repulsion must be the ``same'', 
with the obvious difference of spin-flavour overlaps, as the physics of the 
$KN$ exotic scattering so that we must be able to derive a common origin for
 these two experimental facts. We were also able to derive 
 reasonable $\pi N$ and $\pi \Delta$ couplings for the very same strength of the 
quark 
potential which gave good $KN$ phase shifts, and all this evidence putted 
together allow us to hope to obtain, at least at a qualitative level, a 
unified picture for low energy hadronic reactions including a microscopically 
derived $NN$ potential in terms of one single parameter, if the quark 
masses are neglected. 

In section II we briefly 
describe the model we will be using. The masses of the deltas and nucleons, 
together with the respective Salpeter amplitudes will be obtained in section 
III. In section IV we discuss the set up of the RGM equations appropriated 
for the $NN$ scattering. Section V is devoted to the evaluation of particular 
overlaps needed to solve the $NN$ RGM equation and finally in section VI we 
present and discuss the results.  We conclude with some final remarks.

\section{QUARK MODEL WITH CHIRAL SYMMETRY BREAKING}

\subsection{Generator of the Condensed Vacuum}
In this model we assume a new vacuum $|{\tilde 0}\rangle$ generated from 
the trivial vacuum $|0\rangle$, through the condensation of 
$q{\bar q}$ pairs
\begin{equation}
|{\tilde 0}\rangle=e^{Q_{0}^{\dagger}-Q_{0}}\:\: |0\rangle
\end{equation}
where the generator of the condensed vacuum can be written as
\begin{equation}
Q^{\dagger}_{0}(\Phi)=\int\:d^{3}p\: \Phi(p)\:M_{ss'}(\theta,\phi)\:
b^{\dagger}_{fcs}({\bf p}) d^{\dagger}_{fcs'}(-{\bf p}).
\end{equation}
$b^{\dagger}$ and $d^{\dagger}$ are the creation and annihilation
operators of $q$ and ${\bar q}$, respectively.
$\Phi(p)$ is the radial term of the $q{\bar q}$ wave function and
\begin{equation}\label{angular}
M_{s_{1}s_{2}}(\theta,\phi)=-\sqrt{8\pi}\: \sum_{m_{L},m_{s}}
(1m_{L}\: 1m_{S}|00)
(\frac{1}{2}s_{1}\: \frac{1}{2}s_{2}|1m_{S})\: Y_{1m_{L}}(\theta,\phi) .
\end{equation}
stands for the angular term coupled with the spin. 
The vacuum can be understood in terms of a coherent superposition
of pairs $q{\bar q}$
with the vacuum $J^{PC}$ quantum numbers $O^{++}$ and thus can only
have $^{S}L_{J}$ equal to $^{3}P_{0}$. This coupling is contained in
(\ref{angular}).
The condensed vacuum $|{\tilde 0}\rangle$ obtained is orthogonal to 
$| 0\rangle$ if $\Phi(p)$ differs from zero.

The Fock space quark and antiquark annihilators $b$ and $d$, 
carry indices for flavour, spin and color.
In the new vacuum $|\tilde{0} \rangle$, they are related to the old
ones by a Valatin-Bogoliubov transformation conserving the usual commutation
relations,
\begin{equation}
\left\{ b^{\dagger}_{fsc}({\bf p}),b_{f's'c'}({\bf p}')\right\}=
\left\{ d^{\dagger}_{fsc}({\bf p}),d_{f's'c'}({\bf p}')\right\}=
\delta_{ff'}\delta_{ss'}\delta_{cc'}({\bf p}-{\bf p}') .
\end{equation}

\subsection{Rotation of Spinors}
The field operator $\psi_{fc}({\bf x})$ is defined as
\begin{equation}
\psi_{fc}({\bf x})=\int{\frac{d^{3}p}{(2\pi)^{3/2}}
\left[
u_{s}({\bf p})b_{fsc}({\bf p})+v_{s}({\bf p})d^{\dagger}_{fsc}(-{\bf p})
\right]
e^{i\: {\bf p}.{\bf x}}
}
\end{equation}
Summation over repeated indices is assumed. 
In order to satisfy the formal invariance of $\psi$ under the 
Valatin-Bogoliubov transformation,
the rotation of operators in the Fock-space must be compensated with an 
inverse rotation of spinors $u$ and $v$ and we obtain
\begin{eqnarray}
u_{s}({\bf p})&=&\frac{1}{\sqrt{2}}\left[
\sqrt{1+\sin\varphi(p)}+\sqrt{1-\sin\varphi(p)}\: \hat{{\bf p}}.\vec{\alpha}
\right]\: u^{0}_{s},  \nonumber \\
v_{s}({\bf p})&=&\frac{1}{\sqrt{2}}\left[
\sqrt{1+\sin\varphi(p)}-\sqrt{1-\sin\varphi(p)}\: \hat{{\bf p}}.\vec{\alpha}
\right]\: v^{0}_{s} \label{spinors}  
\end{eqnarray}
where we define the chiral angle
\begin{equation}
\varphi(p)=arctan\left(\frac{m}{p}\right)+2\Phi(p)
\end{equation}
With this expression, where $m$ is the current mass of the quark, 
it is easy to verify that for massless quarks the condensation of vacuum 
generates, through the Valatin-Bogoliubov transformation, an effective  mass.

\subsection{Normal Ordered Hamiltonian}
Once the vacuum has been rotated, it is convenient to normal order
the second quantization Hamiltonian of the model, 
which is
\begin{equation}\label{hamiltonean}
\hat{H}=\int{d^{3}x[\hat{H}_{0}({\bf x})+\hat{H}_{I}({\bf x})]},
\end{equation}
where $\hat{H}_{0}$ is the Hamiltonian density of the Dirac field,
\begin{equation}
\hat{H}_{0}({\bf x})=\psi^{\dagger}({\bf x})
\left(m_{q}\beta-i\vec{\alpha}.\vec{\nabla} \right)\: \psi({\bf x}),
\end{equation}
and $\hat{H}_{I}$ an effective instantaneous interaction term
\begin{equation}
\hat{H}_{I}({\bf x})=\frac{1}{2}\int d^{3}y\: 
\bar{\psi}({\bf x})\gamma^{0}\frac{\lambda^{a}}{2}\psi({\bf x})\:
\:V({\bf x}-{\bf y})\:
\: \bar{\psi}({\bf y})\gamma^{0}
\frac{\lambda^{a}}{2}\psi({\bf y})
.
\end{equation}
The $\lambda^{a}$'s are the Gell-Mann color matrices and $m_{q}$
is the quark mass. $\psi$ carries indices for flavour and color.
Flavour is conserved in both ${\hat H}_{0}$ and ${\hat H}_{I}$.

The normal ordering relatively to the new vacuum gives us the Hamiltonian
in the form
\begin{equation}
\hat{H}={H}_{0}+\hat{H}_{2}+\hat{H}_{2}^{A}+\hat{H}_{4}
\end{equation}
where the subscript is the number of operators in each term.
$H_{0}$ represents the energy of the new vacuum.
\begin{equation}
H_{0}=\:N_{f}\:W\int {d^{3}p \over (2\pi)^3}
\left[
2C({\bf p})
-3\left(E({\bf p})+m_{q}\:\sin\varphi(p)+p\:\cos\varphi(p)
\right)
\right]
\end{equation}
where $N_{f}$ is the number of flavours,
\begin{eqnarray}\label{quarkenergy}
W&=&\int \frac{d^3 x}{(2 \pi )^3} \ , 
\ V({\bf p})= \int \frac{d^3 x}{(2 \pi )^3} V({\bf x})\; 
e^{i {\bf p}\cdot {\bf x}},
\nonumber \\
C({\bf p})&=&\int d^{3}p'\: V({\bf p}-{\bf p'}), \nonumber \\
E({\bf p})&=&\sin\varphi(p) \: A({\bf p})+\cos\varphi(p)\: B({\bf p}), \nonumber \\
A({\bf p})&=&m_{q}+\frac{2}{3}\int 
d^{3}k\: V({\bf p}-{\bf k})\: \sin\varphi (p),   \nonumber \\
B({\bf p})&=&p+\frac{2}{3}\int d^{3}k\: V({\bf p}-{\bf k})\: \cos\varphi (p)\: 
{\bf \hat{p}.\hat{k}}   
\end{eqnarray}

It is possible to show that this energy turns out to be 
smaller than the energy of the trivial ``empty'' vacuum ($ |0\rangle$;
where $ \varphi=0$). The operator 
$\hat{H}_{2}$ is given by
\begin{equation}
\hat{H_{2}}=\int d^{3}p\: E(p)\:
\left[ b^{\dagger}_{fsc}({\bf p})\: b_{fsc}({\bf p})+ d^{\dagger}_{fsc}
(-{\bf p})\: d_{fsc}(-{\bf p})  \right]
 \end{equation},
where $E(p)$ is the energy of a free quark.

The anomalous Bogoliubov term  has the form
\begin{eqnarray}
{\hat H}_{2}^{A}&=&\int d^{3}p\: 
\left[ A(p)\: \sin\varphi (p)-B(p)\: \cos\varphi (p) \right] \nonumber \\
&&.\left[ M_{ss'}b^{\dagger}_{fsc}({\bf p})\: d^{\dagger}_{fs'c}(-{\bf p})+
 M^{*}_{ss'}d_{fsc}(-{\bf p})\: d_{fs'c}({\bf p})  \right].
\end{eqnarray}
It has the form of the vacuum generator $Q_{0}$
an can creates $q{\bar q}$ pairs on the vacuum.
It is now important to remember that the quiral angle, which define
the physical vacuum, has not yet be chosen. 
To achieve this we can say that $\varphi(p)$ minimize the energy of the
new vacuum. With this condition we get 
\begin{equation}\label{mass-gap}
A({\bf p})\: \cos\varphi(p)-B({\bf p})\: \sin\varphi(p)=0
\end{equation}
and $simultaneously$ we get the anomalous Bogoliubov to vanish.
Expression (\ref{mass-gap}) is called {\it mass-gap equation}
and defines the chiral angle $\varphi(p)$ with an obvious functional 
dependence on the chosen potential. An alternative approach 
\cite{VilLiu}, at least at the BCS level, 
is to use the Ward identity for the ${\hat H}_{4}$ kernel
in the derivation of the mass-gap equation.

The term ${\hat H}_{4}$ 
\begin{equation}
{\hat H}_{4}=\frac{1}{2}\int d^{3}p d^{3}k d^{3}q\: V({\bf q})\:
\left( \frac{\lambda^{a}_{c_{1}c_{2}}\lambda^{a}_{c_{1}c_{2}}}{4}\right)\:
\sum_{j,l=1}^{4} :\Theta_{c_{1}c_{2}}^{j}({\bf p},{\bf p}+{\bf q})\:
\Theta_{c_{3}c_{4}}^{l}({\bf k},{\bf k}-{\bf q}) :
\end{equation}
gives 16 terms grouped in ten different
four-quark amplitudes. Such amplitudes are combinations of  the
following four vertices:
\begin{eqnarray}
&&\Theta^{1}_{c'c}({\bf p},{\bf p'})\equiv u^{\dagger}_{s'}({\bf p'})
u_{s}({\bf p})\: b^{\dagger}_{fs'c'}({\bf p'})b_{fsc}({\bf p}), \nonumber \\
&&\Theta^{2}_{c'c}({\bf p},{\bf p'})\equiv -v^{\dagger}_{s'}({\bf p'})
v_{s}({\bf p})\: d^{\dagger}_{fsc}(-{\bf p})d_{fs'c'}(-{\bf p'}), \nonumber \\
&&\Theta^{3}_{c'c}({\bf p},{\bf p'})\equiv u^{\dagger}_{s'}({\bf p'})
v_{s}({\bf p})\: d^{\dagger}_{fs'c'}({\bf p'})d^{\dagger}_{fsc}(-{\bf p}), 
\nonumber \\
&&\Theta^{4}_{c'c}({\bf p},{\bf p'})\equiv v^{\dagger}_{s'}({\bf p'})
u_{s}({\bf p})\: d^{\dagger}_{fs'c'}(-{\bf p'})b_{fsc}({\bf p}).
\end{eqnarray}

\subsection{Spontaneous Chiral Symmetry Breaking}
It was shown \cite{KN} that chiral symmetry is spontaneously broken
in this model. We briefly explain the idea. For massless quarks, 
the Hamiltonian (\ref{hamiltonean}) is invariant  under the global
chiral transformation
\begin{equation}
\psi \rightarrow e^{-i\alpha^{a}T^{a}\gamma_{5}}\:\psi
\end{equation}
where $T^{a}$ are the $SU(N_{f})$ generators and the corresponding
axial charges are,
\begin{equation}
Q_{5}^{a}=\int d^{3}x\: {\bar \psi}\: \gamma_{0}\gamma_{5}T^{a}\: \psi .
\end{equation}

The Bogoliubov transformation on the Fock space of 
the creation and annihilation operators can be cast as a rotation through 
an angle $\Phi=\varphi /2$ and this fact, together with the invariance of 
the field operator $\psi$ under these transformations produces a corresponding 
counter-rotation on the spinorial basis. This is enough for the $Q^{a}_{5}$
to acquire, for non vanishing $\varphi(p)$, an anomalous term which is
the composite operator for the pion creation and thus furnishing 
the only nonzero contribution for $Q^{a}_{5}|{\tilde 0}\rangle$.
At the BCS level, $Q^{a}_{5}|{\tilde 0}\rangle \not= 0$ implies that the pion
mass must be zero and the solution of the corresponding Salpeter equation
for the pion confirms this fact. Furthermore  we obtain, as expected, that 
the pion relative wave function is given by $\sin\varphi(p)$.

\subsection{Harmonic Potential}
 In what concerns the phenomenological potential $V$, 
both the harmonic and linear confining potentials 
have been studied \cite{VilLiu},
together with different combinations of Dirac vertices.  
In the present work we will use the 
Coulomb harmonic confining potential,
\begin{equation}\label{potential}
V({\bf x}-{\bf y})=-\frac{3}{4}K^{3}_{0}\:({\bf x}-{\bf y})^{2}\; , \
V({\bf q})=\frac{3}{4}\:K^{3}_{0}\: \Delta_{\bf q}\: \delta({\bf q})
\end{equation}
which furnished the best results in previous calculations.

Solving Eqs. (\ref{quarkenergy}) and 
(\ref{mass-gap}) both with this potential, we can obtain the explicit 
expression for the energy of a free quark,
\begin{equation}
E_{k}=m_{q}\: \sin\varphi+k\: cos\varphi-\frac{(\varphi ')^{2}}{2}
-\frac{\cos^{2}\varphi}{k^{2}}  
\end{equation}
as well as the mass-gap equation,
\begin{equation}\label{massgapharm}
p^{2}\: \varphi ''+2p\varphi '+2m_{q}p^{2}\cos\varphi-
2p^{3}\sin\varphi + \sin 2\varphi = 0, 
\end{equation}
where units of $K_{0}=1$ are used from now on.

The mass-gap equation (\ref{massgapharm}) was solved 
in Ref \cite{Yaouanc,Modelo},
with the boundary 
conditions
\begin{equation}
\varphi(0)=\frac{\pi}{2},\:\:\: \varphi(p\rightarrow\infty)=0.
\end{equation}
The chiral angle has been studied in detail 
but we point out that in the limit of zero current quark mass and zero
potential, which forces the chiral angle to be zero, expression
(\ref{spinors}) yields the massless Dirac spinors. The other trivial limit 
arises for very massive fermions with $\varphi=\frac{\pi}{2}$, for all small
momenta. In this case and for this region of momentum, 
Eq. (\ref{spinors}) becomes an identity.

\section{NUCLEON AND $\Delta$ BOUND-STATES}

Bound-state equation for mesons and baryons were already studied in
Ref. \cite{Yaouanc,Modelo,VilKrein}. Here we report on more precise results
for the  hadronic radii and 
masses, both for the nucleon and the $\Delta$.

\subsection{The Nucleon Wave Function}

The nucleon 
internal wave function $\phi_{N}({\bf p}_{1},{\bf p}_{2},{\bf p}_{3})$,
where ${\bf p}_{i}$ is the momentum of quark $i$, 
includes the internal degrees of freedom of the  three quarks as
color, spin and isospin.
In the moment space, it is usual to use the ground state 
harmonic oscillator wave function 
\begin{equation}\label{nucleon}
\phi_{000}({\bf p}_{i})=\left(\frac{1}{\pi \alpha^{2}}\right)^{3/4}\:
e^{-\frac{{p}_{i}^{2}}{2\alpha^{2}}}
\end{equation}
as a variational trial wave function for the baryon Salpeter equation, with 
a variational parameter $\alpha$. It was already shown \cite{KN} that it 
provides an excellent approximation to the solution of 
Eq. (\ref{BarSal}) both for the nucleon and the $\Delta$.

Both internal wave functions are orbitally symmetric, $SU(4)$ spin-isospin
symmetric and $SU(3)$ color singlets.
For instance, the  wave function of a nucleon is
\begin{equation}
\phi_{N}=\phi_{000}({\bf p}_{1})\: \phi_{000}({\bf p}_{2})\: \phi_{000}
({\bf p}_{3})
\: \frac{1}{\sqrt 2}(D_{s}D_{f}+F_{s}F_{f})\:\: \epsilon
\end{equation}
where D (F) is a three quark state which is symmetric (antisymmetric) in the
labels of the first and second quarks:
\begin{eqnarray}
D^{\uparrow}_{s}&=&\frac{1}{\sqrt{6}}\left(
\downarrow\uparrow\uparrow + \uparrow\downarrow\uparrow 
-2\uparrow\uparrow\downarrow 
\right)
\nonumber \\
F^{\uparrow}_{s}&=&\frac{1}{\sqrt{2}}\left(
\downarrow\uparrow\uparrow -\uparrow\downarrow\uparrow 
\right)
\end{eqnarray}
and $D_{f}$ and $F_{f}$ are equivalent expressions provided we make the 
substitutions $\uparrow \rightarrow u,\: \downarrow \rightarrow d$ 
in the case of protons and 
$\uparrow \rightarrow u,\: \downarrow \rightarrow  -d$ in the case 
of the neutrons.
 The usual normalized color component is totally antisymmetric,
\begin{equation}
\epsilon=\frac{1}{\sqrt{6}}\: \epsilon_{ijk}\: |ijk\rangle.
\end{equation}

\subsection{Baryons Bound State Equation}

For the present study the only quantity needed is the radius of the nucleon 
in order to solve NN scattering. Therefore we solve the bound state equation
neglecting the negative-energy component, since it corresponds
to channels of much higher energy.

Since our Hamiltonian is instantaneous and there are no negative-energy
channels, the bound-state equation can be written as
\begin{equation}
H\:|\psi\rangle = M \: |\psi\rangle,
\end{equation}
where $|\psi\rangle$ is an eigenstate of the Hamiltonian with mass M.

As in Ref. \cite{KN,fpindelta,VilKrein} we will restrict our calculations 
to bare baryons i.e.
hadrons without coupled hadronic channels, 
in the simplest S-wave configuration,
with quarks of the same current mass, 
leading to only one chiral angle $\varphi$,
the same dispersion relation $E(p)$ for the three quarks and only one set of 
spinors $u$ and $v$ in the vertices $\Theta$. 

We begin with the creation operator for baryons which has the form
\begin{eqnarray}
\Psi^{\dagger}_{B}&=&\int\: d^{9}p
\: \delta({\bf p}_{1}+{\bf p}_{2}+{\bf p}_{3})\:
\psi({\bf p}_{1},{\bf p}_{2},{\bf p}_{3}) \nonumber \\
&&
\epsilon_{c_{1}c_{2}c_{3}}\:
\chi_{f_{1}f_{2}f_{3}s_{1}s_{2}s_{3}}\:
b^{\dagger}_{f_{1}s_{1}c_{1}}({\bf p}_{1})
b^{\dagger}_{f_{2}s_{2}c_{2}}({\bf p}_{2})
b^{\dagger}_{f_{3}s_{3}c_{3}}({\bf p}_{3})
\end{eqnarray}
where $\psi$, $\chi$ and $\epsilon$ represent, respectively
the momentum, the spin-flavour and the color components of the
baryon wave function. 
Calculating $(\hat{H}_{2}+\hat{H}_{4})\Psi_{B}^{\dagger}\: |0\rangle$
we have for the baryon bound state equation:
\begin{eqnarray}\label{BarSal}
&&[M-3E(p_{1})]\: \chi_{s_{1}s_{2}s_{3}}\:
\psi({\bf p}_{1},{\bf p}_{2},{\bf p}_{3})\:
=-2\int \: d^{3}q \: V({\bf q})\: 
\nonumber  \\
&&
[u^{\dagger}_{s_{1}}({\bf p}_{1})\:
u_{s_{3}}({\bf p}_{1}-{\bf q})\:
u^{\dagger}_{s_{2}}({\bf p}_{2})\:
u_{s_{4}}({\bf p}_{2}+{\bf q})]\:
\chi_{s_{4}s_{5}s_{6}}\:
\psi({\bf p}_{1}-{\bf q},{\bf p}_{2}+{\bf q},{\bf p}_{3})
\end{eqnarray}
with ${\bf p}_{1}+{\bf p}_{2}+{\bf p}_{3}=0$, for any flavour indices in the 
function $\chi$. The factor $-2$ includes a contribution of $-2/3$ from
the contraction of the color matrices times the number of quark pairs.
Thus, the bound-state equation (\ref{BarSal}) has the form of 
a Schr\"odinger equation
with an effective interaction
\begin{equation}
V=-\frac{2}{3}\int \: d^{3}q \: V({\bf q})\:
[u^{\dagger}_{s_{1}}({\bf p}_{1})\:
u_{s_{3}}({\bf p}_{1}-{\bf q})\:
u^{\dagger}_{s_{2}}({\bf p}_{2})\:
u_{s_{4}}({\bf p}_{2}+{\bf q})]\:
\chi_{s_{4}s_{5}s_{6}}\:
\psi({\bf p}_{1}-{\bf q},{\bf p}_{2}+{\bf q},{\bf p}_{3}).
\end{equation}

Calculations made with the harmonic potential (\ref{potential}) give
\begin{equation}
V_{q_{1}q_{2}}=V_{q_{1}q_{2}}^{SI}+V_{q_{1}q_{2}}^{SS}+V_{q_{1}q_{2}}^{T}
+V_{q_{1}q_{2}}^{SO}
\end{equation}
including the spin independent,  hyperfine,  tensor and
spin-orbit interactions \cite{VilKrein}.

\subsection{Nucleon and $\Delta$ masses}

In order to check the consistence of our model, we calculate the $N$ and
$\Delta$ masses.
In the approximation of $N$ and $\Delta$ with only one S-wave channel, the 
spin-orbit and tensorial terms vanish and the remainder bound-state 
equation is
\begin{eqnarray}
&&\left\{
3E(p_{1})-M-\frac{3}{2}\nabla^{2}_{{\bf p}_{12}}
+\frac{3}{4}(\varphi '(p_{1}))^{2}\right.
+\frac{3\left[1-\sin\varphi(p_{1})\right]}{p_{1}^{2}}  \nonumber \\
&+&\left.\left[\frac{3}{4}-\frac{1}{3}S(S+1)\right]
\left[1-\sin\varphi(p_{1})\right]\left[1-\sin\varphi(p_{2})\right]
\frac{{\bf {\hat p}}_{1}.{\bf {\hat p}}_{2}}{p_{1}p_{2}}
\right\}\: \psi({\bf p}_{1},{\bf p}_{2},{\bf p}_{3})=0,
\end{eqnarray}
where $S$ is the total spin and $\nabla^{2}_{{\bf p}_{12}}$ stands for the
Laplacian with respect to the relative momentum $({\bf p}_{1}-{\bf p}_{2})/2$.
The mass of the baryon is then given by
\begin{eqnarray}
M&=&\langle
3E(p_{1})-M-\frac{3}{2}\nabla^{2}_{{\bf p}_{12}}
+\frac{3}{4}(\varphi ')^{2}(p_{1})
+\frac{3\left[1-\sin\varphi(p_{1})\right]}{p_{1}^{2}} 
\rangle_{{\bf p}_{1}+{\bf p}_{2}+{\bf p}_{3}=0} \nonumber \\
&&
+\left[\frac{3}{4}-\frac{1}{3}S(S+1)\right]
\langle\left[1-\sin\varphi(p_{1})\right]\left[1-\sin\varphi(p_{2})\right]
\frac{{\bf {\hat p}}_{1}.{\bf {\hat p}}_{2}}{p_{1}p_{2}}
\rangle_{{\bf p}_{1}+{\bf p}_{2}+{\bf p}_{3}=0}
\end{eqnarray}
where the following notation has been introduced:
\begin{equation}
\langle
f({\bf p}_{1},{\bf p}_{2},{\bf p}_{3})
\rangle_{{\bf p}_{1}+{\bf p}_{2}+{\bf p}_{3}=0}\equiv
\int d^{9}p\: \delta({\bf p}_{1}+{\bf p}_{2}+{\bf p}_{3})
\: f({\bf p}_{1},{\bf p}_{2},{\bf p}_{3})
\: \left|
\psi({\bf p}_{1},{\bf p}_{2},{\bf p}_{3})
\right|^{2} 
\end{equation}

Since we do not have an analytic expression for the chiral angle $\varphi$,
the calculations were done with the good approximation 
\begin{equation}
\sin\varphi(p)=e^{-p^{2}/(2\beta^{2})}
\end{equation}
where $\beta=0.582$.

Figure \ref{figure1} shows the results for the $N \Delta$ masses, in units
of $K_{0}$,
as a function of the variational parameter $\alpha$, for the case of
massless current quarks. The results are presented in table \ref{table1}
considering the harmonic potential with a strength constant $247$ MeV.

The bare nucleon mass obtained gives a mean radius of $0.66$ fm,
which is reduced when the strength of potential is increased.
As an example, $K_{0}=290$ MeV gives $0.57$ fm.
The introduction of finite current quark masses almost does not change
the results we have obtained.

The difference between the bare nucleon mass and its physical counterpart
is a consequence of the coupling to $N\pi$ channels which 
is expected to lower the nucleon bare mass somewhere between ($-300$) and 
($-400$) MeV \cite{Thomas}.
The effect of coupled channels should also contributes to the physical 
$N-\Delta$ difference ($100-200$ MeV).

\section{RGM EQUATION FOR NN SCATTERING}

It is well known from potential scattering theory
that a free spherical wave converging towards the origin is almost 
unaffected by a potentials which range is
of the order of  $\sqrt{L(L+1)}/k$.  The repulsive short range
potential between two nucleons has almost no effect on waves for which
$\sqrt{L(L+1)}/k>0.8$.
This fact  motivates the choice of the relative S wave
to test the repulsive core of  NN scattering also revealed by experimental 
low-energy scattering phase shifts in $^{1}S_{0}$ and $^{3}S_{1}$ partial
waves.

To study the central microscopic NN repulsion it suffices to consider 
the hyperfine quark-quark potential $V_{q_{1}q_{2}}^{SS}$,
\begin{equation}\label{potss}
V_{q_{i}q_{j}}^{SS}=
\frac{\lambda_{i}^{a}}{2}.\frac{\lambda_{j}^{a}}{2}
\frac{\left[1-\sin\varphi(p_{i})\right]\:\left[1-\sin\varphi(p_{j})\right]}
{p_{i}p_{j}}
\: {\bf {\hat p}}_{i}.{\bf {\hat p}}_{j}\: {\bf S}_{i}.{\bf S}_{j}.
\end{equation}
As we are neglecting the pionic cloud responsible for the long and
intermediate range forces, 
it is obvious that we will not be able to reproduce the NN phase shifts.
However, the results will allow us to show how much the quark structure 
of the nucleons influences the NN S-wave scattering.

To investigate the Nucleon-Nucleon scattering we make use of the Resonating
Group Method (RGM) introduced by Wheeler \cite{Wheeler}, 
and profusely applied by several groups \cite{Ribeiro,Oka,Warke,Faessler}.
In this method the two clusters relative motion
wave function is obtained by requiring it to minimize energy of the system.
Then, we must to minimize the matrix element
\begin{equation}
I=\langle N_{1}\: N_{2}| (\hat{H}-E)|
N_{1}\: N_{2}\rangle 
\end{equation}
where $\hat{H}=\hat{H}_{2}+\hat{H}_{4}$ is the Hamiltonian in the
second quantization formalism ones, 
$E$ represents the total energy of the system and $|N_{1}\: N_{2}\rangle$
describes the six quarks system clustered in nucleons $N_{1}$ and $N_{2}$.

The Pauli principle is taken account by demanding the antisymmetry of the 
total six quarks ensemble which arises naturally
through the Wick ordering of quark fields.
Since the wave functions of nucleons are already
antisymmetrized, we obtain,  
\begin{eqnarray}
\langle N_{1}N_{2}|(\hat{H}-E)|N_{1} N_{2}\rangle &=&
\langle N_{1}N_{2}|(H-E)\:36(1-9P_{14})(1-{\cal P})
|N_{1} N_{2}\rangle \nonumber \\
&\equiv&
\langle N_{1}N_{2}|(H-E)\:{\cal A}|N_{1} N_{2}\rangle
\end{eqnarray}
where $P_{ij}$ stands for the permutation of the quarks $i\in N_{1}$ 
and $j\in N_{2}$ and
${\cal P}=P_{14}P_{25}P_{36}$ gives the permutation of the nucleons $1$ 
and $2$. ${\cal A}$ is the usual designation of the antisymmetrizer 
of six quarks in two nucleons. The operator H is defined 
in first quantization.

At this stage it is interesting  
to see that the factor $(1-{\cal P})$
implies, in a natural fashion, 
the usual antisymmetry condition at the baryon level. 
In fact, $(1-{\cal P})$ acts on the system of two nucleons and makes
the matrix element vanish unless the condition $L+S+T=odd$ holds.

The Jacobi coordinates adapted to this problem are
\begin{eqnarray}\label{JacCoor}
&&{\bf p}_{\rho}=\frac{1}{\sqrt{2}}({\bf p}_{1}-{\bf p}_{2}) ,\:\:\: \:\:\:
{\bf p}_{\lambda}=\frac{1}{\sqrt{6}}({\bf p}_{1}+{\bf p}_{2}-2{\bf p}_{3})
\nonumber  \\* 
&&{\bf p}_{\xi}=\frac{1}{\sqrt{2}}({\bf p}_{4}-{\bf p}_{5}),
\:\:\: \:\:\:
{\bf p}_{\gamma}=\frac{1}{\sqrt{6}}({\bf p}_{4}+{\bf p}_{5}-2{\bf p}_{6})
\nonumber \\*
&&{\bf p}_{r}=\frac{1}{\sqrt{6}}({\bf p}_{1}+{\bf p}_{2}+{\bf p}_{3}) 
-\frac{1}{\sqrt{6}}({\bf p}_{4}+{\bf p}_{5}+{\bf p}_{6})
\nonumber \\* 
&&{\bf p}_{cm}=\frac{1}{\sqrt{6}}({\bf p}_{1}+{\bf p}_{2}+{\bf p}_{3}+
{\bf p}_{4}+{\bf p}_{5}+{\bf p}_{6})  .
\end{eqnarray}
As a result of this choice, each nucleon is described by one of
the two sets $\left\{(\rho,\lambda)\right\}$, $\left\{(\xi,\gamma)\right\}$ 
of coordinates. The relative motion is described through ${\bf p}_{r}$.

The Hamiltonian in first quantization, in the center of mass frame, becomes
\begin{equation}
H=T_{\rho}+T_{\lambda}
+T_{\xi}+T_{\gamma}+T_{r}+\sum_{i<j}V_{ij}
\end{equation}
The subscripts $\rho , \ \lambda, \ \xi, \gamma, \ r$ and  
$cm$ correspond to the Jacobi 
coordinates of equation (\ref{JacCoor}) and $V_{ij}$ is the potential of 
interaction  between the quarks $i$ and $j$. 
Using a more explicit notation,
\begin{equation}
I=36\langle \chi ({\bf p_{r}})|\langle    
\phi^{1}(\xi_{1})\phi^{2}(\xi_{2})
|(H-E)(1-9P_{14})
(1-{\cal P})|\phi^{1}(\xi_{1})\phi^{2}(\xi_{2})\rangle 
|\chi ({\bf p_{r}}) \rangle_{{\bf p}_{cm}={\bf 0}}
\end{equation} 
where the arguments $\xi_{1}$ and $\xi_{2}$ denote the sets 
of internal degrees of freedom in the 
wave function $\phi^{1,2}$ of each nucleon,
including  ${{\bf p}_\rho,{\bf p}_\lambda,{\bf p}_\xi,{\bf p}_\gamma}$ 
in the momenta space, 
and $\chi({\bf p}_{r})$
is the wave function for the relative motion of the two nucleons.

Expanding $\chi({\bf p}_{r})$
in spherical harmonics, 
\begin{equation}\label{chiexpans}
\chi({\bf p}_{r})=\sum_{m}\: \sum_{l}\: \chi_{lm}({\bf p}_{r}),
\end{equation}
where $\chi_{lm}({\bf p}_{r})$ includes both radial and orbital components,
we obtain that
for the particular case of S relative wave function  
$\chi_{S}({\bf p}_{r})=\chi_{S}(-{\bf p}_{r})$, and we can drop 
the permutation ${\cal P}$ between the two nucleons, 
providing the product 
$\phi^{1}(\xi_{1})\:\phi^{2}(\xi_{2})$ of the two wave functions
is antisymmetrized. 
The RGM  equation is obtained minimizing the matrix 
element $I$ with respect to $\chi$,
\begin{equation}
\frac{\partial I}{\partial \langle \chi_{S} |}=0\Leftrightarrow
\langle \phi^{1}(\xi_{1})\phi^{2}(\xi_{2})|(H-E)(1-9P_{14})
|[\phi^{1}(\xi_{1})\phi^{2}(\xi_{2})]
\rangle 
|\chi_{S} \rangle =0.
\end{equation}

It is easy to verify that the specific dependence of color in
$V_{ij}$ yields  a null matrix element when the permutation is not present. 
The nucleon mass $M_{N}$ is defined by
\begin{equation}
H_{N}|N_{1}\rangle=(T_{\rho}+T_{\lambda}+V_{12}
+V_{13}+V_{23})|N\rangle
=M_{N}|N_{2}\rangle ,
\end{equation} 
with an analogous expression for the second nucleon. 

Now we introduce the eigenfunctions  ${\phi_{nlm}^{\alpha}}\equiv {\phi_{a}}$ 
of the harmonic oscillator with parameter $\alpha$ 
as a complete set for the expansion of 
the relative wave function$|\chi_{S}\rangle$
\begin{equation}
|\chi_{S}\rangle=\sum_{ab}|\phi_{a}\rangle\langle\phi_{a}| |\phi_{b}\rangle
\langle \phi_{b}|\chi_{S}\rangle . \label{expansion}
\end{equation}
Using the fact that 
$E_{r}=E-2\: M_{N}$, we get
\begin{equation}\label{rgm}
\left[(T_{r}-E_{r})(1+\sigma_{ab}|\phi_{a}\rangle\langle\phi_{b}|)
+v_{ab}|\phi_{a}\rangle\langle\phi_{b}|\right]\: |\chi_{S}\rangle=0 ,
\end{equation}
where
\begin{eqnarray}
\sigma_{ab}&=&-9\langle\phi_{a}({\bf p}_r)|\langle 
\phi^{1}(\xi_{1})\phi^{2}(\xi_{2})|P_{14}|
[\phi^{1}(\xi_{1}),\phi^{2}(\xi_{2})]
\rangle|\phi_{b}({\bf p}_r)\rangle \equiv -9  \langle P_{14}\rangle ,
\nonumber \\
v_{ab}&=&-9\langle\phi_{a}({\bf p}_r)|\langle 
\phi^{1}(\xi_{1})\phi^{2}(\xi_{2})|\sum_{i \in N_{1}, 
j \in N_{2}} V_{ij}P_{14}|[\phi^{1}(\xi_{1}),\phi^{2}(\xi_{2})]
\rangle|\phi_{b}({\bf p}_r)\rangle  .
\end{eqnarray}
We have to note that Eq. (\ref{expansion}),
associated to expansion (\ref{chiexpans}), implies that 
$\phi_{b}=\phi_{n,0,0}$.
We will retain $a=b=(0,0,0)$ as an approximation that is 
reasonable due to the fact that the overlap kernel being compact, 
happens to filter out higher wave excitations quite effectively.

Now, if we use
\begin{equation}\label{WavFunc}
|\psi\rangle =\left(1+\sigma_{00}|\phi_{000}\rangle\langle\phi_{000}|\right)
|\chi\rangle
\end{equation}
where $\phi_{000}$ is the wave function of the tridimensional
harmonic oscillator with the relative Jacobi momenta ${\bf p}_{r}$
it is possible to rewrite Eq. (\ref{rgm}) as a non-local Schr\"odinger
type equation:
\begin{equation}\label{EqSchro}
\left(T_{r}-E_{r}+\frac{v_{00}}{1+\sigma_{00}}|\phi_{000}\rangle
\langle\phi_{000}|\right)|\psi\rangle =0
\end{equation}
Of course, wave function (\ref{WavFunc}) gives the same phase shifts
than $|\chi\rangle$.

\section{EVALUATION OF $\sigma$ AND v}

The nine diagrams we have to consider are given in Fig. \ref{figure2}.
The values of $\sigma_{00}$ for the color and spin-flavour spaces were  
directly calculated numerically, performing all the permutations
involved.
To obtain the value of $v_{00}$ in the same spaces, 
we must consider the nine graphs 
of Fig. \ref{figure2}. 
In order to simplify calculations, 
the interactions of color and spin 
can be commuted in permutations given by relations:
\begin{eqnarray}
\frac{\lambda_{i}^{a}}{2}.\frac{\lambda_{j}^{a}}{2}&=&
\frac{1}{2} (P_{ij}^{color}+2\delta_{ij}-\frac{1}{3}) \nonumber \\
\vec{S}_{i}.\vec{S}_{j}&=&
\frac{1}{2} (P_{ij}^{spin}-\frac{1}{2}) 
\end{eqnarray}

The evaluations in momenta space could be made with the appropriate 
graphical rules \cite{Ribeiro2}.
But in our order of approximation, we do not need to use them.
For instance, to obtain $\langle P_{14}\rangle$,
we just have to note that $P_{14}$ does not change the momenta part
of the wave function which describes 
six quarks grouped in two nucleons and with relative wave function
$\phi_{000}$. This result is instantaneously obtained by inspection of
expression 
\begin{equation}
(\phi_{N1}\phi_{N2})_{mom}=
\phi_{000}({\bf p}_{\rho})\: \phi_{000}({\bf p}_{\lambda})
\phi_{000}({\bf p}_{\xi})\: \phi_{000}({\bf p}_{\gamma})
\phi_{000}({\bf p}_{r})
\end{equation}
Therefore $\sigma=1$.

For each one of the nine terms of $v_{00}$, 
the procedure is the same we describe.
We start observing that for the reason given above, permutation $P_{14}$
can be removed. In momenta space, each term of 
$v_{00}$ reveals to be proportional to
\begin{equation}
{\cal V}_{00} \equiv \langle\phi_{000}({\bf p}_r)|\langle 
\phi^{1}(\xi_{1})\phi^{2}(\xi_{2})|(V_{ij})_{mom}
|\phi^{1}(\xi_{1})\phi^{2}(\xi_{2})
\rangle|\phi_{000}({\bf p}_r)\rangle_{{\bf p}_{cm}=0}  ,
\end{equation}
where $(V_{ij})_{mom}$ is the momentum dependent part of the hyperfine 
quark potential.

A numerical evaluation gives ${\cal V}_{00}=-0.0671 K_{0}$.
The calculations were done for the cases $(T,S)=(1,0)$ and
$(T,S)=(0,1)$ and the results are in table \ref{table2},
where the evaluations in the spin-flavour space must be multiplied
by a factor $(1/18)^{2}$.

For instance, we have, for (T,S)=(1,0):
\begin{eqnarray}
\sigma_{00}&=&
 \langle P_{14}\rangle_{c}. \langle P_{14}\rangle_{s\times f}.
\langle P_{14}\rangle_{mom.} =\frac{1}{9} \nonumber \\
v_{00}&=& 
-9\: \sum_{i \in N_{1}, 
j\in N_{2}}\left(\langle V_{ij}P_{14}\rangle_{c}. \langle V_{ij}P_{14}
\rangle_{s\times f}.
\langle V_{ij}P_{14}\rangle_{mom.}\right) 
=-\frac{5}{3} {\cal V}_{00}
\end{eqnarray}
These results are summarized in table \ref{table3}.

\section{RESULTS AND DISCUSSION}

The value of $K_0$, the unique parameter of this model,  was fixed to
$K_{0}=247$ MeV in order to obtain good KN phase shifts.  
This value also gave good $f_{N\pi N}$ and $f_{N\pi\Delta}$  and
was used successfully by the Orsay quark group \cite{Yaouanc} .
We obtain $\alpha=296.4$ MeV and $-{\cal V}_{00}=16.6$ MeV. 
In the most conventional quark models, the oscillator parameter has values of
$250$ MeV \cite{harvey} to $410-420$ MeV \cite{Faessler,Koniuk}.

Since the potential is separable, the phase shifts were obtained 
analytically 
\begin{equation}
\delta(z)=- atan 
\frac{ 4\:  z\: \sqrt{\pi}\: exp(-z^{2})}{\frac{\alpha^{2}}{\lambda m}+4-
8z\: exp(-z^{2})\: \int_{0}^{z}e^{t^{2}}\: dt}
\end{equation}
with
\begin{equation}
\lambda=\frac{v_{00}}{1+\sigma_{00}},\:\:\:
z=\frac{p_{r}}{\alpha}, \:\:\:
E_{r}=\frac{\left(\frac{\sqrt{6}}{2}p_{r}\right)^{2}}{2m}
\end{equation}
where $m$ is the physical reduced mass of the system of two nucleons
and $E_{r}$ stands for the energy of the relative motion in the center
of mass frame. The factor $\sqrt{6}/2$ is needed because $p_r$ is a
Jacobi coordinate.
The  phase shifts obtained
are presented 
in Figs. \ref{figure3} for the $(T,S)=(1,0);(0,1)$ channels.

In order to derive an
equivalent local potential $U_{ELP}$ defined, according to Ref. \cite{Oka}, as
\begin{equation}
\left[
\frac{\hbar^{2}}{2\mu}\nabla^{2}+E_{r}-U_{ELP}(r)
\right]
\psi(r)=0\: ,
\end{equation}
the nonlocal Schr\"odinger equation (\ref{EqSchro}) was solved.
This local potential happens to be energy-dependent.
The numerical accuracy  was tested by reproducing numerically the
previous phase shifts.

It is expected in a microscopic calculation of 
the $NN$ interaction, to obtain for the
short range part a non-local description,
besides the usual meson-exchange picture which is also natural in 
this framework.  There are a few realistic potentials that fit in this
description, with a different height for the core, describing very 
well the two body data. The latter does not seem to be very sensitive to
the height of the core. The Nijmegen potential was chosen for
comparison because it has a version, the NijmI \cite{Nijmegen}, 
with a very soft core, comparable to ours, and has an
optimal $\chi^{2}$ per datum for $np$ data. 
In Fig \ref{figure4}, both our equivalent local potential $U_{ELP}$ and 
the  NijmI potential are depicted. 
It is well known that they should differ in the medium range, due to the
meson exchange contributions which are present in any realistic potential, 
and that we have not yet included.
Nevertheless it is remarkable that, contrary to past experiences,
such a soft repulsive core can be accomodated to reproduce
experimental phase shifts.

Almost all the $NN$ potentials obtained in the framework of quark models
have
comparable range values, but they differ largely in the magnitudes.
For example, Faessler {\it et al} and Oka \cite{Oka,Faessler}
find potentials with peak values between $600$ and $1300$ MeV.
Ribeiro \cite{Ribeiro} built a square potential of magnitude $500$ MeV 
that was approximately phase shift equivalent 
to the ones obtained solving the RGM equation.
The Liberman potential \cite{Liberman} has a peak of $450$ MeV.
Of course, a  direct comparison of the potentials is not absolutely
correct because the methods used to derive each potential 
are quite differents. In order to avoid this problem, we can 
compare the equivalent local potentials obtained by Suzuki and 
Hecht\cite{Suzuki}, 
who used a common method for the models we have referred above.
However we can accept this large range of results if we take in account 
the small relative weight of the potential near the origin.

Although it is usually expected that quark exchange processes 
give the most relevant
contribution to the $NN$ repulsion,
$\omega$ and $\rho$ can still give contributions to the 
short range potential. These mesons, and the other possible ones
required to describe the 
intermediate and long range attraction of the nuclear force,
can be introduced in a consistent way within this model.

\section{conclusion}

In this work we have derived, from a chiral invariant quark model,
the short range part of the $NN$ interaction. There is only one
free parameter in this model, which was already fixed in previous
calculations.
A non-local potential was then derived, with a soft core,
consistent with the recent NijmI realistic 
interaction, that fully describe the experimental $NN$ data.
This is part of a wider program to describe the full nucleon-nucleon
interaction within a quark model.

Finally we find remarkable that
we are able to understand, with
the same parameter, the $NN$ repulsion
and, {\it at the same time},
to reproduce the meson and baryon spectra, 
the vector meson decays, the $f_{N \pi N}$
coupling and to obtain a reasonable $KN$ repulsion.


\begin{figure}
\label{figure1}
\caption{Masses of $N$ and $\Delta$ as a function of the variational
parameter $\alpha$. Both are in units of $K_{0}=1$.}
\end{figure}

\begin{figure}
\label{figure2}
\caption{The nine diagrams that contribute to the potential overlap.}
\end{figure}

\begin{figure}
\label{figure3}
\caption{Phase shifts of $NN$ scattering for $(T,S)=(1,0)$ 
and $(T,S)=(0,1)$ channels.}
\end{figure}

\begin{figure}
\label{figure4}
\caption{Our equivalent local potential (solid line) and the realistic
NijmI potential (dotted line).
The deviation for large distances is due, as usual, to attractive meson
exchange.}
\end{figure}

\begin{table}
\begin{center}
\begin{tabular}{|l|c|c|}
\hline\hline
 & N & $\Delta$  \\
\hline
Mass ($K_{0}$) & 4.88  (1207 MeV) & 5.08 (1255 MeV) \\
\hline
$\alpha$ ($K_{0}$) & 1.20 & 1.21 \\
\hline\hline
\end{tabular}
\end{center}
\label{table1}
\caption{$N, \Delta$ masses and inverse radius $\alpha$ in units of 
$K_{0}=1$. Values in MeV were obtained with $K_{0}$=247 MeV.}
\end{table}

\begin{table}
\begin{center}
\begin{tabular}{|l|c|c|c|c|c|c|}
\hline\hline
 & \multicolumn{3}{c|}{(T,S)=(1,0)} &  \multicolumn{3}{c|}{(T,S)=(0,1)}  \\
\hline
 & color & spin-flavour & mom. & color & spin-flavour & mom. \\
\hline
$\langle P_{14}\rangle$ & $\frac{1}{3}$ & $-12$ & $1$ & $\frac{1}{3}$ 
& $-12$ & $1$ \\
\hline
$\langle V_{14}P_{14}\rangle$ 
& $\frac{4}{9}$ & $93$ & ${\cal V}_{00}$ & $\frac{4}{9}$ 
& $57$ & ${\cal V}_{00}$ \\
\hline
$\langle V_{15} P_{14}\rangle$ 
& $-\frac{2}{9}$ & $-21$ &${\cal V}_{00}$ & $-\frac{2}{9}$ 
& $-21$ &${\cal V}_{00}$ \\
\hline
$\langle V_{16}P_{14}\rangle$ 
& $-\frac{2}{9}$ & $-21$ &${\cal V}_{00}$ & $-\frac{2}{9}$ 
& $-21$ &${\cal V}_{00}$ \\
\hline
$\langle V_{24}P_{14}\rangle$ 
& $-\frac{2}{9}$ & $-21$ &${\cal V}_{00}$ & $-\frac{2}{9}$ 
& $-21$ &${\cal V}_{00}$ \\
\hline
$\langle V_{25}P_{14}\rangle$ 
& $\frac{1}{9}$ & $0$ &${\cal V}_{00}$ & $\frac{1}{9}$ 
& $5$ &${\cal V}_{00}$ \\
\hline
$\langle V_{26}P_{14}\rangle$ 
&$\frac{1}{9}$ & $0$ &${\cal V}_{00}$ & $\frac{1}{9}$ 
& $5$ &${\cal V}_{00}$ \\
\hline
$\langle V_{34}P_{14}\rangle$ 
& $-\frac{2}{9}$ & $-21$ &${\cal V}_{00}$ & $-\frac{2}{9}$ & $-21$ 
&${\cal V}_{00}$ \\
\hline
$\langle V_{35}P_{14}\rangle$ 
& $\frac{1}{9}$ & $0$ &${\cal V}_{00}$ & $\frac{1}{9}$ & $5$ 
&${\cal V}_{00}$ \\
\hline
$\langle V_{36}P_{14}\rangle$ 
& $\frac{1}{9}$ & $0$ &${\cal V}_{00}$ & $\frac{1}{9}$ & $5$ 
&${\cal V}_{00}$ \\
\hline\hline
\end{tabular}
\end{center}
\label{table2}
\caption{Contribution of each diagram in color, spin-flavour and momenta
spaces.}
\end{table} 

\begin{table}
\begin{center}
\begin{tabular}{|l|c|c|}
\hline\hline
 & $(T,S)=(1,0)$ & $(T,S)=(0,1)$  \\
\hline
$\sigma_{00}$ &  $\frac{1}{9}$  & $\frac{1}{9}$ \\
\hline
$v_{00}$ & $-\frac{5}{3}{\cal V}_{00}$ & $-\frac{104}{81}{\cal V}_{00}$ \\
\hline
$\lambda\equiv \frac{v_{00}}{1+\sigma_{00}}$ & 
$-\frac{3}{2}{\cal V}_{00}$ & $-\frac{52}{45}{\cal V}_{00}$ \\
\hline\hline
\end{tabular}
\end{center}
\label{table3}
\caption{Final values for $\sigma_{00}$ and $v_{00}$. 
The relevant quantity to obtain phase shifts is  $\lambda$.}
\end{table}
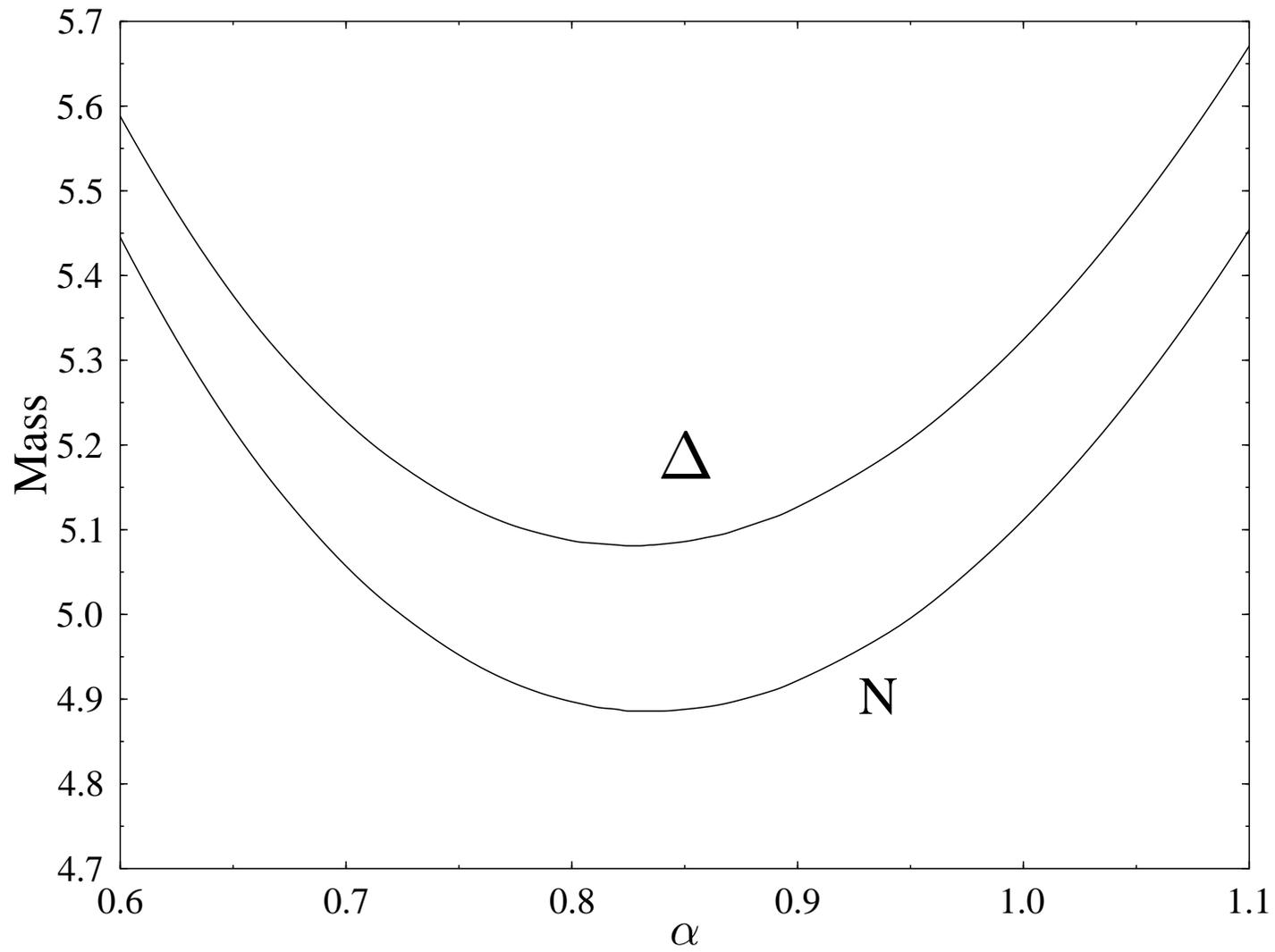
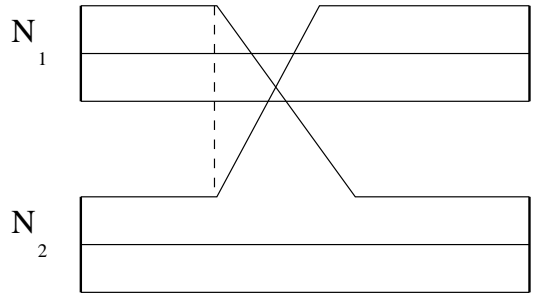
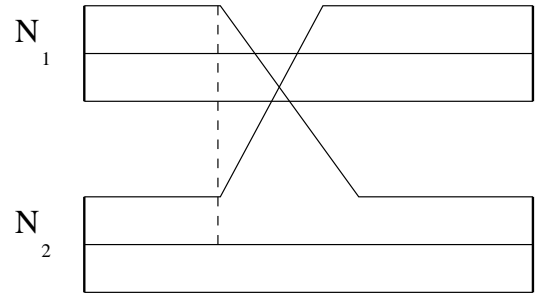
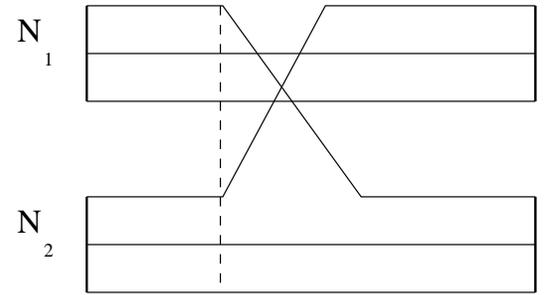
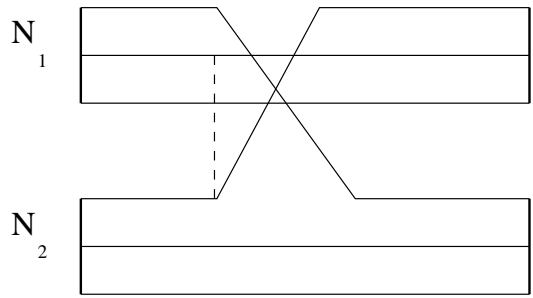
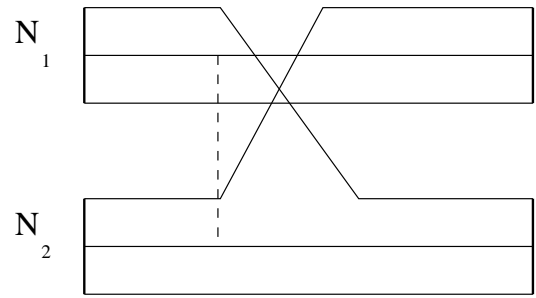
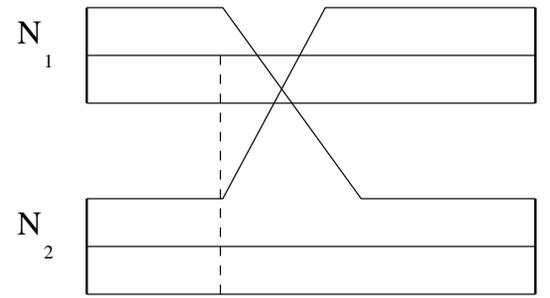
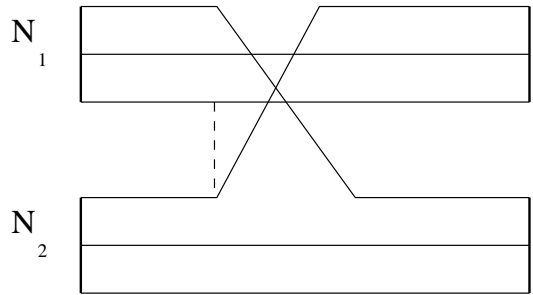
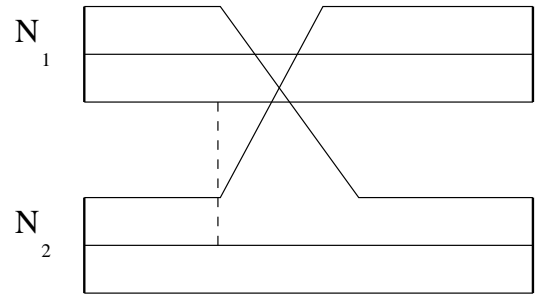
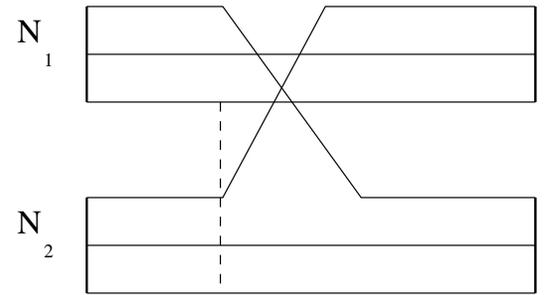
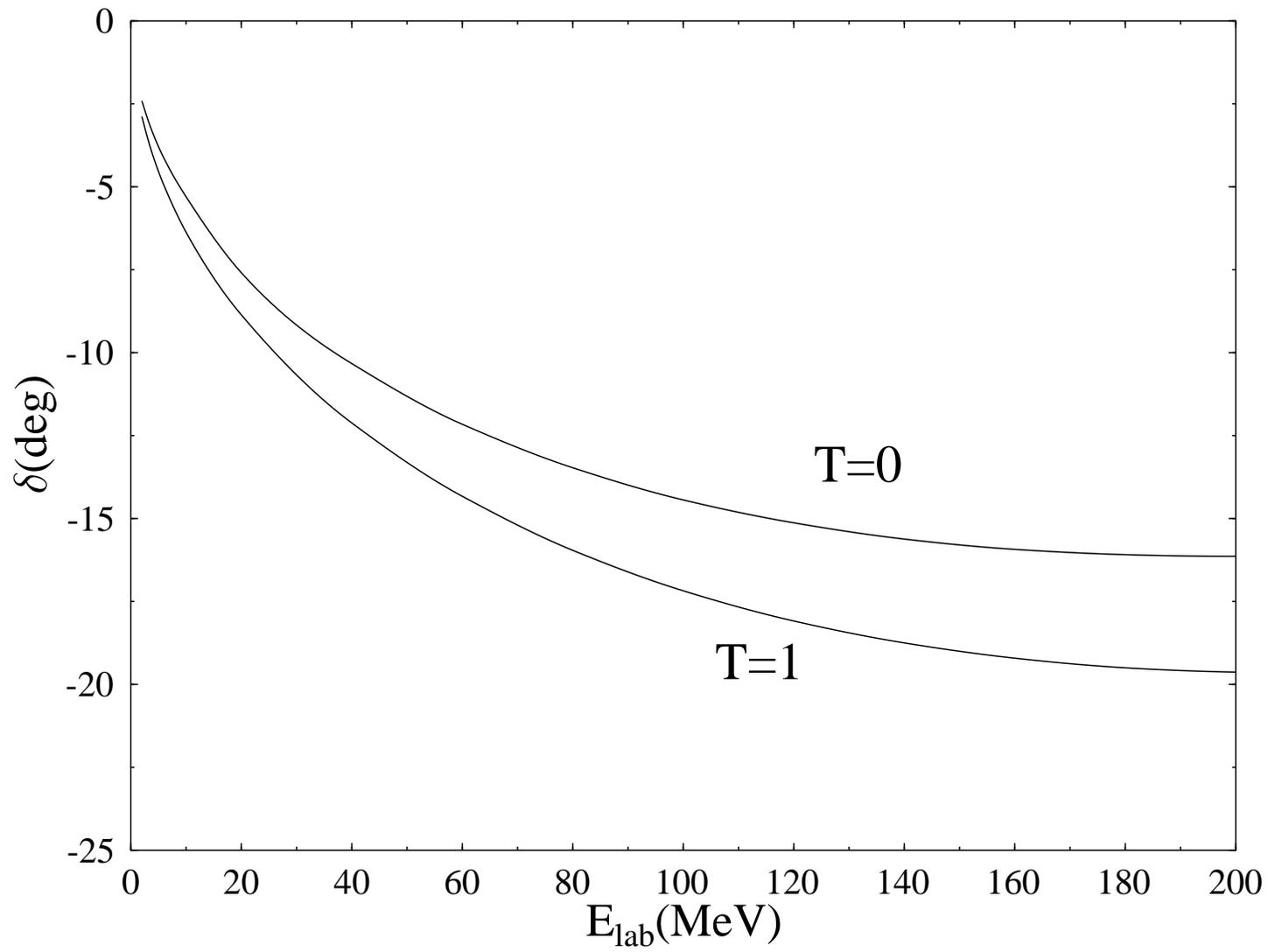
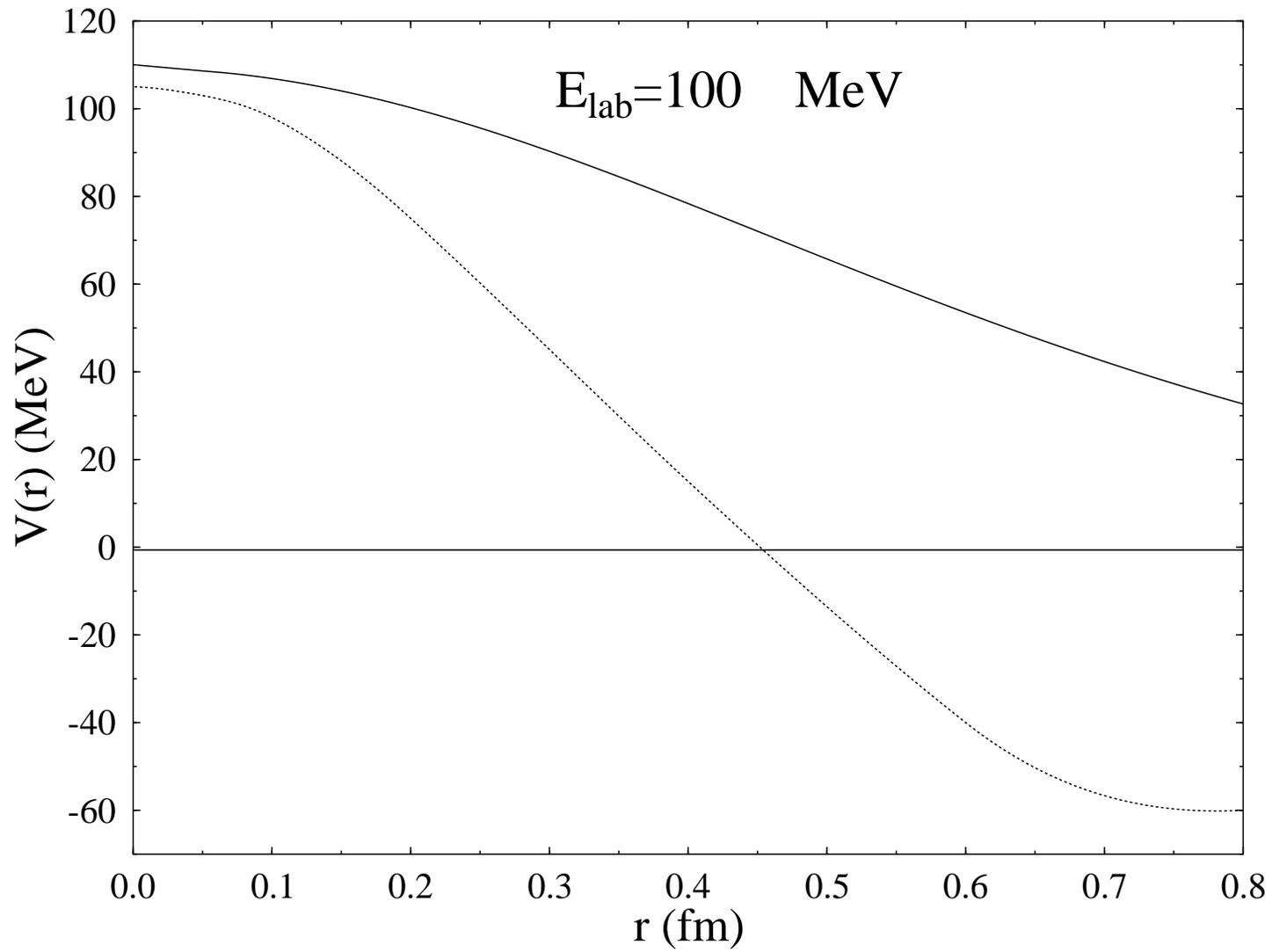
\end{document}